\documentclass[a4paper,11pt]{article}
\pdfoutput=1 

\usepackage{jinstpub}

\usepackage{lineno}
\usepackage[LGRgreek]{mathastext}

\title{Performance assessment of the HERD calorimeter with a photo-diode read-out system for high-energy electron beams}

\author[a,b]{O. Adriani,}
\author[c]{G. Ambrosi,}
\author[d]{M. Antonelli,}
\author[e]{Y. Bai,}
\author[e]{X. Bai,}
\author[f]{T. Bao,}
\author[c]{M. Barbanera,}
\author[a,b]{E. Berti,}
\author[a,b,1]{P. Betti,\note{Corresponding author.}}
\author[g,h]{G. Bigongiari,}
\author[a,b]{M. Bongi,}
\author[d]{V. Bonvicini,}
\author[b]{S. Bottai,}
\author[i,l]{I. Cagnoli,}
\author[e]{W. Cao,}
\author[m]{J. Casaus,}
\author[n,o]{D. Cerasole,}
\author[e]{Z. Chen,}
\author[f]{X. Cui,}
\author[a,b]{R. D'Alessandro,}
\author[o]{L. Di Venere,}
\author[m]{C. Diaz,}
\author[f]{Y. Dong,}
\author[b]{S. Detti,}
\author[c]{M. Duranti,}
\author[o]{F. Gargano,}
\author[e]{J. Gao,}
\author[f]{S. Guo,}
\author[m]{F. Giovacchini,}
\author[p,b]{N. Finetti,}
\author[q]{V. Formato,}
\author[r,c]{Y. Jiang,}
\author[e]{X. Liang,}
\author[e]{R. Li,}
\author[f]{C. Liao,}
\author[f]{X. Liu,}
\author[e]{L. Lyu,}
\author[m]{J. Marin,}
\author[m]{G. Martinez,}
\author[b]{N. Mori,}
\author[s]{A. Oliva,}
\author[b,2]{L. Pacini,\note{Corresponding author.}}
\author[b]{P. Papini,}
\author[o]{R. Pillera,}
\author[d]{C. Pizzolotto,}
\author[f]{Z Quan,}
\author[e]{J.J. Qin,}
\author[i,l]{L. Silveri,}
\author[c]{G. Silvestre,}
\author[e]{D. Shi,}
\author[o]{D. Serini,}
\author[b]{O. Starodubtsev,}
\author[f]{X. Tang,}
\author[b]{A. Tiberio,}
\author[b]{E. Vannuccini,}
\author[m]{M. Velasco,}
\author[e]{B. Wang,}
\author[f]{J. Wang,}
\author[f]{R. Wang,}
\author[f]{Z. Wang,}
\author[f]{M. Xu,}
\author[f]{X. Yang,}
\author[d]{G. Zampa,}
\author[d]{N. Zampa,}
\author[f]{S. Zhang,}
\author[e]{and J. Zheng}

\affiliation[a]{Department of Physics and Astronomy, University of Florence, I-50019 Sesto Fiorentino, Florence, Italy}
\affiliation[b]{INFN sezione di Firenze, I-50019 Sesto Fiorentino, Florence, Italy} 
\affiliation[c]{INFN Sezione Perugia, Istituto Nazionale di Fisica Nucleare, Sezione di Perugia, 06100 Perugia, Italy}
\affiliation[d]{INFN Sezione di Trieste, Padriciano 99, I-34149 Trieste, Italy}
\affiliation[e]{Xi’an Institute of Optics and Precision Mechanics, Chinese Academy of Sciences, 710119, Xi’an, China}
\affiliation[f]{Institute of High Energy Physics, Chinese Academy of Sciences, 100049, Beĳing, China}
\affiliation[g]{Department of Physical Sciences, Earth and Environment, University of Siena, I-53100 Siena, Italy}
\affiliation[h]{INFN Pisa,	Largo B. Pontecorvo, 3 - 56127 Pisa, Italy}
\affiliation[i]{Gran Sasso Science Institute (GSSI), Viale Crispi 7, I-67100 L'Aquila , Italy}
\affiliation[l]{INFN Laboratori Nazionali del Gran Sasso, Via Acitelli 22, I-67100 Assergi, L'Aquila, Italy}
\affiliation[m]{Centro de Investigaciones Energ\'{e}ticas, Medioambientales y Tecnol\'{o}gicas (CIEMAT), E-28040 Madrid, Spain}
\affiliation[n]{Dipartimento Interateneo di Fisica "M.Merlin" dell'Università e del Politecnico di Bari, I-70126, Bari, Italy}
\affiliation[o]{Istituto Nazionale di Fisica Nucleare (INFN)–Sezione di Bari, I-70126 Bari, Italy}
\affiliation[p]{Department of Physical and Chemical Sciences, University of L'Aquila, Via Vetoio, Coppito, 67100 L'Aquila, Italy}
\affiliation[q]{INFN Sezione Roma TorVergata, Istituto Nazionale di Fisica Nucleare, Sezione di Roma Tor Vergata, 00133 Roma, Italy}
\affiliation[r]{Università degli Studi di Perugia, Università di Perugia, 06100 Perugia, Italy}
\affiliation[s]{INFN Sezione Bologna, Istituto Nazionale di Fisica Nucleare, Sezione di Bologna, 40126 Bologna, Italy}

\emailAdd{pietro.betti@fi.infn.it}
\emailAdd{lorenzo.pacini@fi.infn.it}

\abstract{The measurement of cosmic rays at energies exceeding $100\ TeV$ per nucleon is crucial for enhancing the understanding of high-energy particle propagation and acceleration models in the Galaxy. HERD is a space-borne calorimetric experiment that aims to extend the current direct measurements of cosmic rays to unexplored energies. The payload is scheduled to be installed on the Chinese Space Station in 2027. The primary peculiarity of the instrument is its capability to measure particles coming from all directions, with the main detector being a deep, homogeneous, 3D calorimeter. The active elements are read out using two independent systems: one based on wavelength shifter fibers coupled to CMOS cameras, and the other based on photo-diodes read-out with custom front-end electronics. A large calorimeter prototype was tested in 2023 during an extensive beam test campaign at CERN. In this paper, the performance of the calorimeter for high-energy electron beams, as obtained from the photo-diode system data, is presented. The prototype demonstrated excellent performance, e.g., an energy resolution better than $1\%$ for electrons at $250\ GeV$. A comparison between beam test data and Monte Carlo simulation data is also presented.}

\arxivnumber{????} 

\begin{document}
\maketitle
\flushbottom

\section{The HERD mission.}
\label{sec:HERD}

\begin{figure}[t]
 	\centering 
 	\includegraphics[width=.9\textwidth]{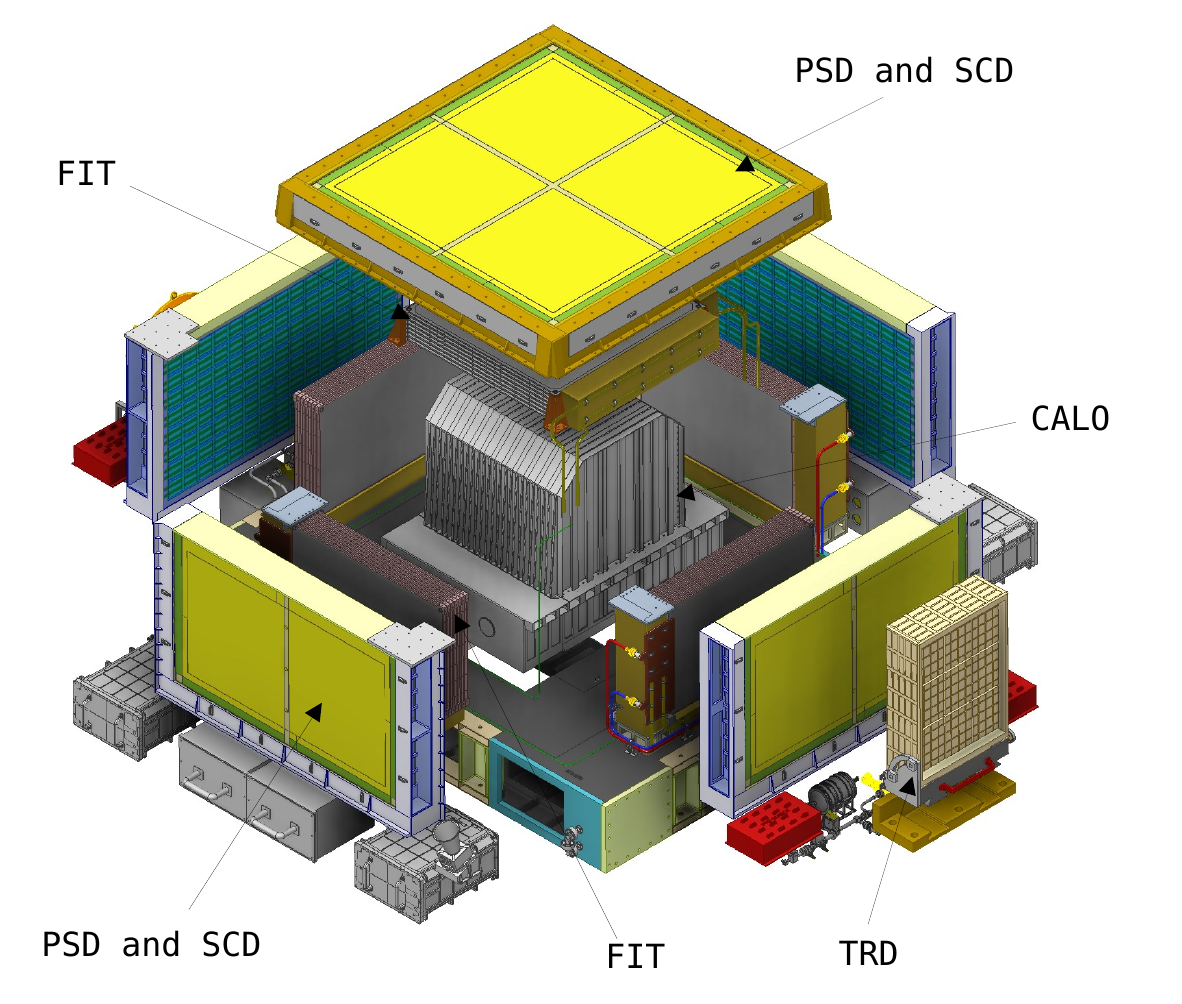}
 	\caption{\label{fig:HERD} Expanded 3D image of HERD: the arrows indicate some of the sub detectors.}
\end{figure}

After the 2000s, a new era of precise cosmic-ray (CR) direct measurements has begun~\cite{bib:CRreview2023}. Balloon experiments, space-borne spectrometers, and calorimeters have conducted numerous observations, leading to the discovery of several unexpected features in CR spectra. These results have significantly enhanced our understanding of CR acceleration and propagation. Consequently, several new theoretical models have been developed, e.g., \cite{bib:Blasi2012}. Even though these experiments provided outstanding information related to CR physics, the energy range of current space experiments is limited. For example, the AMS-02 spectrometer Maximum Detectable Rigidity is a few TV, while the limited acceptance of space calorimeters like CALET~\cite{bib:CALETCalib2017} and DAMPE~\cite{bib:DAMPEDetector2017} allows for the observation of particles up to a few hundred TeV per nucleon.\\
The High Energy cosmic-Radiation Detection (HERD) project~\cite{bib:HERDGargano2021, bib:HERDBetti2024} is a space-based experiment scheduled to be installed on the Chinese Space Station (CSS) around 2027. This initiative involves a partnership among various Chinese and European academic and research institutions. With its innovative design, HERD will push the boundaries of CR spectrum measurements into previously unexplored regions by space-based instruments. The experiment aims to detect protons and nuclei up to the PeV per nucleon region, where a change in the spectral index of the total particle spectrum is observed. Furthermore, HERD will make the first direct observations of the electron+positron flux up to tens of TeV, where various theoretical models predict new spectral features due to nearby CR sources~\cite{bib:ExpectedElectronTeV_2016}. The experiment will also measure the photon spectrum ranging from $500\ \text{MeV}$ to $100\ \text{TeV}$ and will monitor the gamma-ray sky with a wide field of view~\cite{bib:HERDphotons2021}.\\
A large acceptance is required to achieve the HERD goals; thus, the experiment is designed to properly reconstruct particles entering the detector from each face, in contrast to typical CR experiments, which are usually designed like telescopes and accept particles coming only from the top face. This innovative design was first accurately studied by the CaloCube R\&D project~\cite{bib:CaloCubeBottai2017, bib:CaloCubeElectron2021} and then adapted and improved by the HERD collaboration. An image of the tentative design of the detector is shown in Figure \ref{fig:HERD}. HERD consists of several sub-detectors: starting from the outside, Silicon Charge Detectors (SCD)~\cite{bib:HERDSCD_2023}, Plastic Scintillator Detectors (PSD)~\cite{bib:HERDDavide1_2023,bib:HERDDavide2_2023,bib:HERDPSD_2021}, and Fiber Trackers (FIT)~\cite{bib:HERDfit_2021} are present on each detector face, excluding the bottom one. These detectors reconstruct the absolute value of particle charge and particle directions, and allow gamma-ray observation~\cite{bib:HERDphotons2021}. Furthermore, on a single lateral face, a small Transition Radiation Detector (TRD)~\cite{bib:HERDtrd_2020} is present; this provides an independent measurement of TeV proton and GeV electron primary energy.
\\
The core of the instrument is the calorimeter (CALO): it is briefly described in the next section while details about the expected performance of this sub-detector are well described in \cite{bib:HERDPacini2021}.
\section{The calorimeter and the photo-diode read-out system.}
\label{sec:CALO}

\begin{figure}[t]
 	\centering 
 	\includegraphics[width=.81\textwidth]{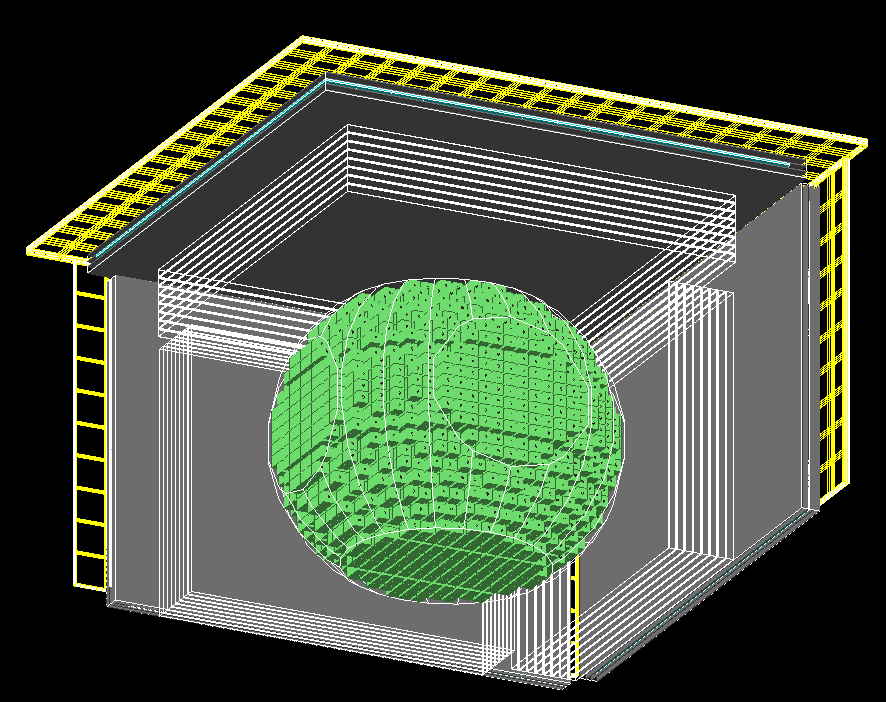}
 	\caption{\label{fig:caloImages} An image of the HERD detector extracted from the GEANT4 simulation. The LYSO crystals are represented with green cubes.}
\end{figure}

\begin{figure}[t]
 	\centering 
 	\includegraphics[width=.99\textwidth]{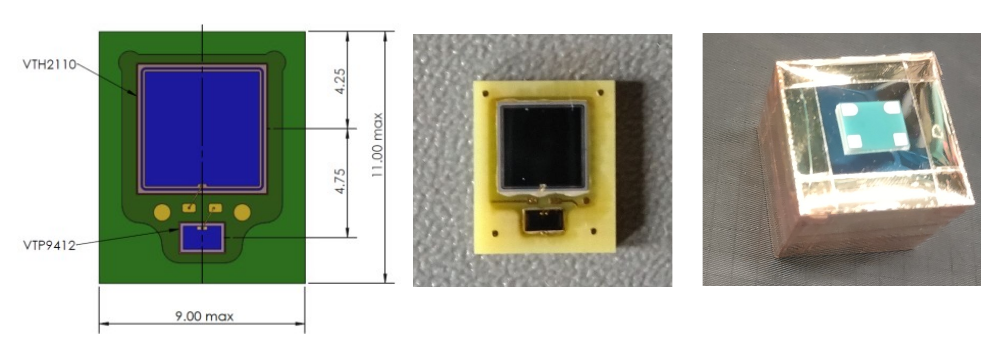}
 	\caption{\label{fig:PDimage} Starting from the left, this figure shows the monolithic package design, a sensor top face picture, and a picture of the sensor glued to a LYSO cube wrapped with an Enhanced Specular Reflector (ESR).}
\end{figure}

The CALO consists of about 7500 cubic scintillating crystals made of LYSO. This material has been selected among other scintillators due to its good light yield and high density, which significantly improve the CALO performance for protons and nuclei. A detailed comparison of calorimeter performance made of different materials is well described in \cite{bib:CaloCubeBottai2017}. The main properties of LYSO are listed in Table~\ref{tab:LYSO}.
\begin{table}[h]
\begin{tabular}{| c | c | c | c | c | c | c |}
	\hline	
	Material & $\rho\ (g/cm^3)$ & $\lambda_I\ (cm)$ & $X_0\ (cm)$ & $\lambda_{max}\ (nm)$ & $\tau\ (ns)$ & $L. Y. (photons/keV)$\\ \hline	
	LYSO & $7.25$ & $21$ & $1.1$ & $\sim425$ &  $40$ & $\sim30$ \\ \hline	
\end{tabular}
\caption{\label{tab:LYSO} Main LYSO crystal parameters: $\rho$ is the density, $\lambda_I$ is the nuclear interaction length, $X_0$ is the radiation length, $\lambda_{max}$ is the emission spectrum peak, $\tau$ is the decay time and $L. Y.$ is the light yield.}
\end{table}
The cubes are geometrically arranged to obtain a pseudo-ellipsoid shape, which increases the geometrical acceptance of the CALO with respect to other configurations, e.g., cubic shape. This design is also compliant with the requirements of the HERD mission: a weight of about $4\ \text{tons}$, a maximum volume of a few $m^3$, and a maximum power consumption of $1.5\ \text{kW}$. An image of the CALO is shown in Figure \ref{fig:caloImages}; the LYSO cubes are colored in green. The total depth of the CALO for vertical particles is about 55 radiation lengths ($X_0$) and about 3 proton interaction lengths ($\lambda_i$). This allows for total absorption of electromagnetic showers up to tens of $TeV$ and a high interaction probability for protons and nuclei.\\
The performance of a slightly different version of the CALO geometry is well described in \cite{bib:HERDPacini2021}; a summary is reported here. The effective geometric factor is better than $2\ \text{m}^2\text{sr}$ for electrons and $1\ \text{m}^2\text{sr}$ for protons and nuclei. The energy resolution at $10\ \text{TeV}$ is about $2\%$ for electrons, $30\%$ for protons, and $22\%$ for carbon nuclei. Thanks to the 3D sampling of the shower, electrons are clearly separated from the proton background: the residual proton contamination is a few percent up to the $10\ \text{TeV}$ region. By reconstructing the shower in the CALO, the shower axis is identified with an angular resolution better than one degree for TeV particles. The updated performance of the CALO with the new geometry is currently under study and will be published in a dedicated paper. Small improvements with respect to the previous estimation are expected.\\
The HERD collaboration is assessing the feasibility of reading out the LYSO cubes with two independent systems: the advantages of having this "double read-out system" are described in \cite{bib:HERDpd2022} and \cite{bib:HERDLiu2023}. In summary, two systems will decrease the systematic errors related to the energy scale calibration. The first system~\cite{bib:HERDLiu2023} is based on Wavelength Shifter Fibers (WLF) connected to Intensified scientific CMOS. The second one is used for acquiring the data discussed in this paper: the details of the system configuration are reported in \cite{bib:HERDpd2022}, while a quick review is provided here. This system consists of photo-diodes (PDs) coupled with custom front-end electronics: this configuration was first studied and validated by the CaloCube collaboration~\cite{bib:CaloCuebHardware2019, bib:CaloCubeElectron2021}. A custom sensor is glued to each LYSO cube. The scintillation light detector consists of two PDs with different active areas assembled in a monolithic package, developed by INFN in collaboration with Excelitas Technologies. The large area PD (LPD) detects small energy deposits in LYSO while the small one (SPD) detects high signals. Table~\ref{tab:PD} includes important parameters of the PDs.
\begin{table} [h]
	\begin{tabular}{| c | c | c | c | c | c |}
		\hline	
		Part number & Active area & Junction cap. & Dark current & Response @ $420\ nm$ & Rise time\\ \hline	
		Large PD & $25\ mm^2$ &  $<30\ pF$ &  $<0.2\ nA$ & $\sim 0.15\ A/W$ & $\sim15\ ns$ \\ \hline	
		Small PD & $1.6\ mm^2$ &  $<6\ pF$ & $<7\ nA$ & $\sim 0.2\ A/W$ & $\sim15\ ns$ \\ \hline	
	\end{tabular}
\caption{\label{tab:PD} PD parameters for a bias voltage of about $50\ V$.}
\end{table}
The SPD active area is covered by an optical filter with a transmission of about $5\%$ at the LYSO wavelength ($\sim 420\ nm$) in order to further increase the LPD/SPD signal ratio: for future prototypes and the CALO flight model, the transmission of the filter will be smaller ($\sim1.5\%$) to further increase the dynamic range. Figure \ref{fig:PDimage} shows some sensor pictures. 
The main component of the front-end electronics is the HiDRA3.0 chip, which is an updated version of the CASIS~\cite{bib:CASIS}. Thanks to the unique design of a double-gain charge-sensitive amplifier, HiDRA features low noise (about 2500 equivalent electrons), low power consumption ($3.75\ \text{mW/channel}$), and a large dynamic range (from a few fC to $52.6\ \text{pC}$). The chip also has self-trigger capabilities and anti-saturation circuitry. Custom kapton cables are used to connect each PD to a dedicated HiDRA channel. The expected properties of this version of the PD read-out system are summarized in Table~\ref{tab:system_performance}.
\begin{table} [h]
\begin{tabular}{| c | c | c | c | c | c| c|}
	\hline	
	mean MIP & Noise & MIP S/N ratio & Filter & LPD/SPD & SPD S/N & Saturation \\ \hline	
	$150\ ADC$ & $<50\ ADC$&  $>3\ ADC$ & $\sim 5\%$ & $\sim 500$ & $>30 $ & $\sim80\ TeV$ \\ \hline	
\end{tabular}
\caption{\label{tab:system_performance} Expected features of the PD read-out system.}
\end{table}
Please note that prototype properties differ with respect to the expected performance of the flight model~\cite{bib:HERDpd2022}. Preliminary validations of the PD read-out system performance are presented in \cite{bib:HERDPietroPD2023, bib:HERDPietroPD2024}.

\subsection{Large prototype: design, simulation and beam test campaign.}
\label{sec:Prototype}

\begin{figure}[t]
 	\centering 
 	\includegraphics[width=.99\textwidth]{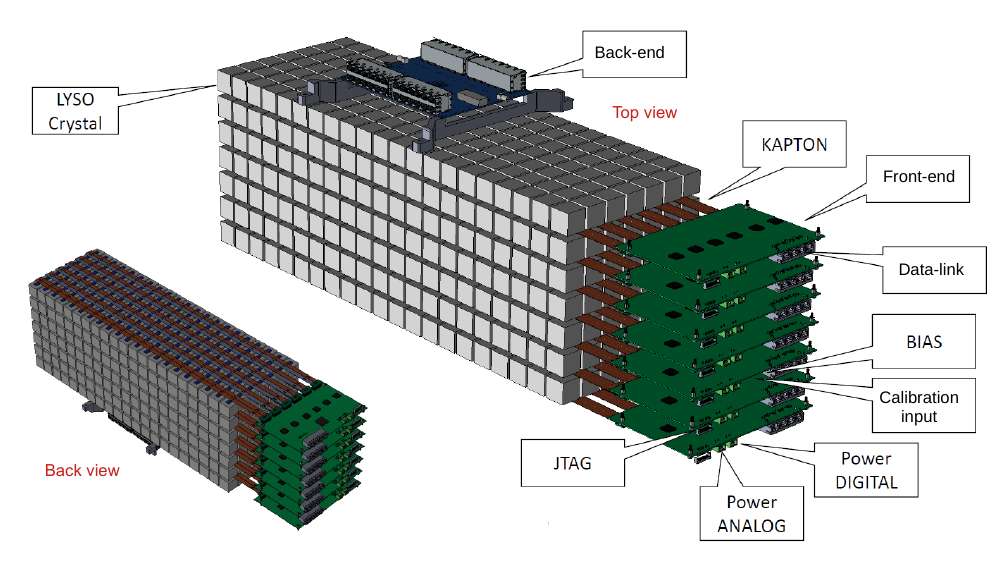}
 	\caption{\label{fig:PrototypeDesign} Mechanical design of the prototype main components, e.g LYSO crystals, back-end board, front-end boards and kapton cables.}
\end{figure}

\begin{figure}[t]
 	\centering 
 	\includegraphics[width=.99\textwidth]{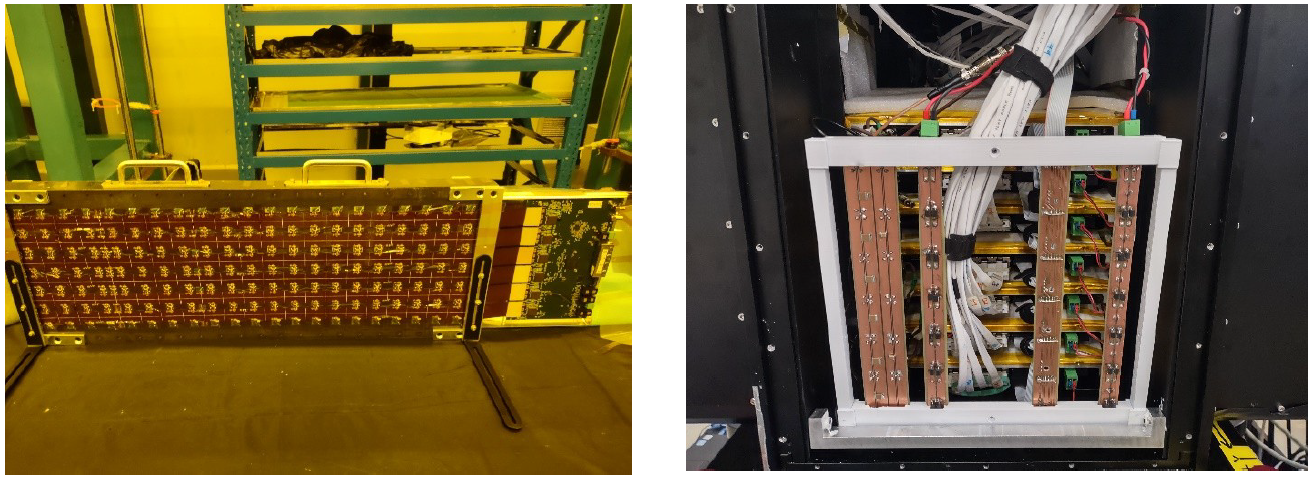}
 	\caption{\label{fig:LayerPhotos} Photographs of the bottom face of a horizontal layer (here disposed in vertical position) and the back of the CALO, where the digital and power cables are connected to front-end boards.}
\end{figure}

\begin{figure}[t]
 	\centering 
    \includegraphics[width=0.7\textwidth]{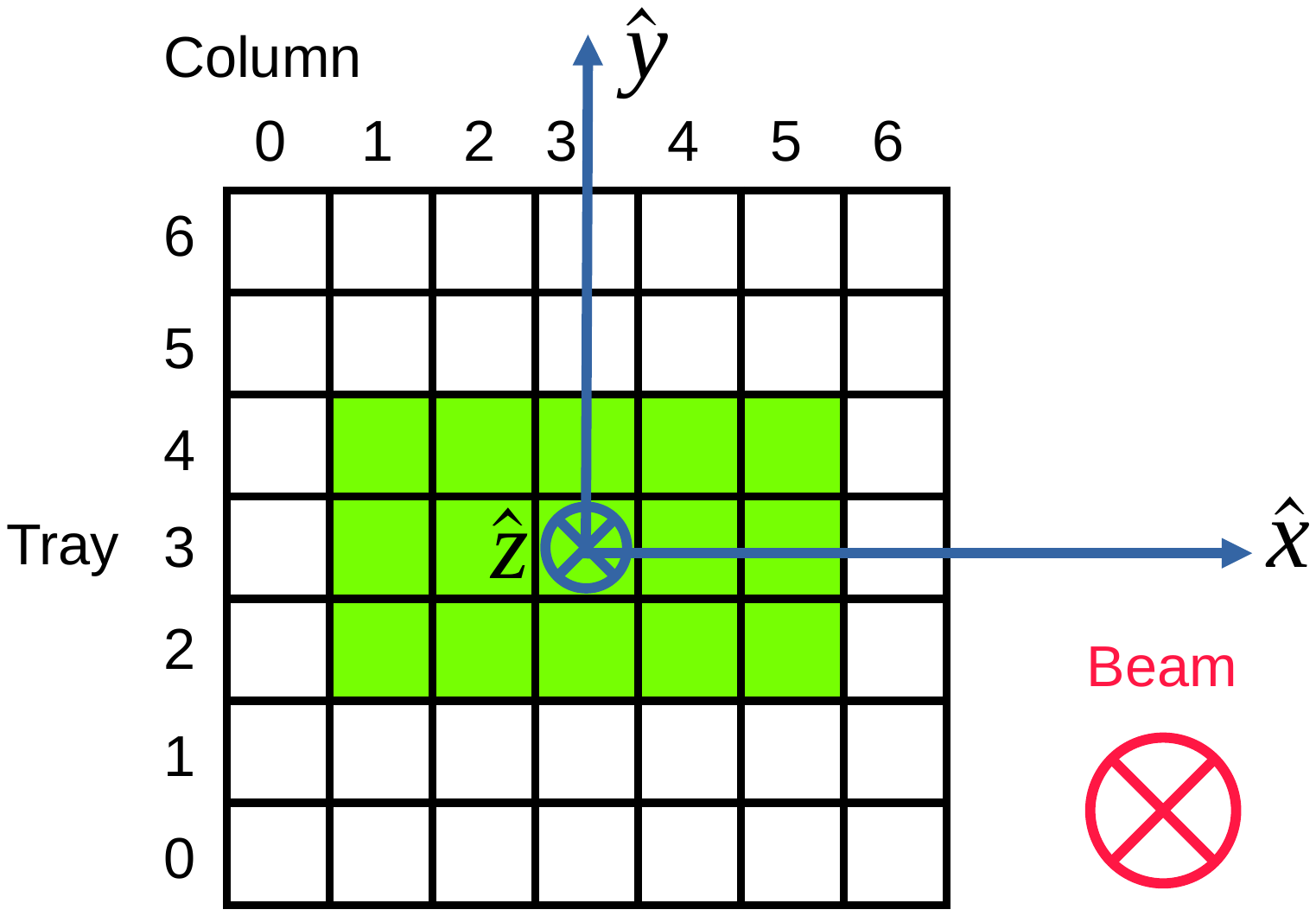}
 	\caption{\label{fig:coordinate} A section of the CALO perpendicular to the beam direction which shows the coordinate system of the CALO. The highlighted green area represents the crystals of every layer considered in this analysis.}
\end{figure}

To evaluate the CALO performance using high-energy beams, a prototype composed of 7x7x21 LYSO crystals was assembled during 2023. This consist of 7 horizontal layers, containing 7x21 cubes. All the crystals are equipped with both read-out systems, i.e., WLF and PDs. Thanks to this prototype, the HERD collaboration was able to conduct an initial test of this unique double read-out system with high-energy particles. The 3D design and two images of the prototype are shown in Figure \ref{fig:PrototypeDesign} and Figure \ref{fig:LayerPhotos}. The PDs are connected to the front-end boards with custom kapton cables accurately designed for the HERD experiment~\cite{bib:HERDpd2022}. The main components of the front-end boards are: 12 HiDRA (version 3) chips, 3 ADC chips, and 1 FPGA. A single front-end board is employed to read-out the PDs of a single horizontal layer; thus, seven boards were used to fully equip the prototype. Finally, a back-end board is assembled on top of the CALO. The main goals of this board are to acquire data from the front-end, handle the fast trigger information, communicate with the main PC, and configure the front-end parameters for the current acquisition run.\\
This prototype and several HERD sub-detectors were tested during an extensive campaign at CERN with PS (east area, EA) and SPS (north area, NA) secondary beams. In this paper, we will present the results obtained with high-energy electrons acquired in the NA. The analysis of muon data acquired in the EA will be briefly described, as it is involved in the calibration procedure of the CALO. Several sub-detectors were placed in front of the CALO, including different prototypes of PSD, SCD, FIT, a trigger box made of plastic scintillator, and an additional charge tagger used for heavy ion tagging. For the analysis described in this paper, data from the SCD made by INFN-Perugia (Italy)~\cite{bib:HERDSCD_2023} were employed to reconstruct the particle tracks; the details of the SCD detector and track reconstruction will be discussed in a separate paper.\\
An accurate Monte Carlo (MC) simulation of the HERD prototypes based on GEANT4 has been developed. The mechanical structure of the CALO is simulated with high precision in order to reproduce the real object. The main approximation in the simulation is the absence of the kapton cables, which, being only a few hundreds of microns thick, should not significantly affect the CALO performance. All the sub-detectors have been included in the MC simulation in order to properly account for the materials placed in front of the CALO. Unfortunately, the accuracy of the simulation of the passive materials of a few sub-detectors is currently limited due to the lack of detailed information about the mechanical structures. To properly emulate the electronic noise and other minor effects of the PD system, the first step in the simulation data analysis involves a digitization procedure of the CALO Monte Carlo data by using of several parameters extracted from the CALO calibration (Section~\ref{sec:calibration}).\\
As a reference system during the analysis, crystals are identified by three coordinates as shown in Figure \ref{fig:coordinate}: tray number (from bottom to top) along the $y$-axis, column number (from left to right) along the $x$-axis, and layer number along the $z$-axis. Specifically, the beam entrance layer is 0, while the last one is 20. The set of all cubes with the same tray and column number is referred to as "line". Unfortunately, due to the strict timeline of the prototype assembly and delays of component procurement, the new front-end boards were produced and immediately installed on the prototype. This has led to some non-negligible consequences: most of the channels of tray 5, column 0 and column 6 have high noise. Therefore, we decided to consider only the central cubes (green area of Figure \ref{fig:coordinate}) in the presented analysis. Given that, this section of the CALO still has more than 2 (3.5) Molière radii of active material vertically (above and below) and horizontally (on the left and on the right of the center), which is enough to contain more than $95\%$ of the deposited energy. In the following sections, the term CALO will refer only to the $5 \times 3 \times 21$ calorimeter core.

\section{Calorimeter calibration with muon beam data.}
\label{sec:calibration}

\begin{figure}[t]
 	\centering 
    \includegraphics[width=0.56\textwidth]{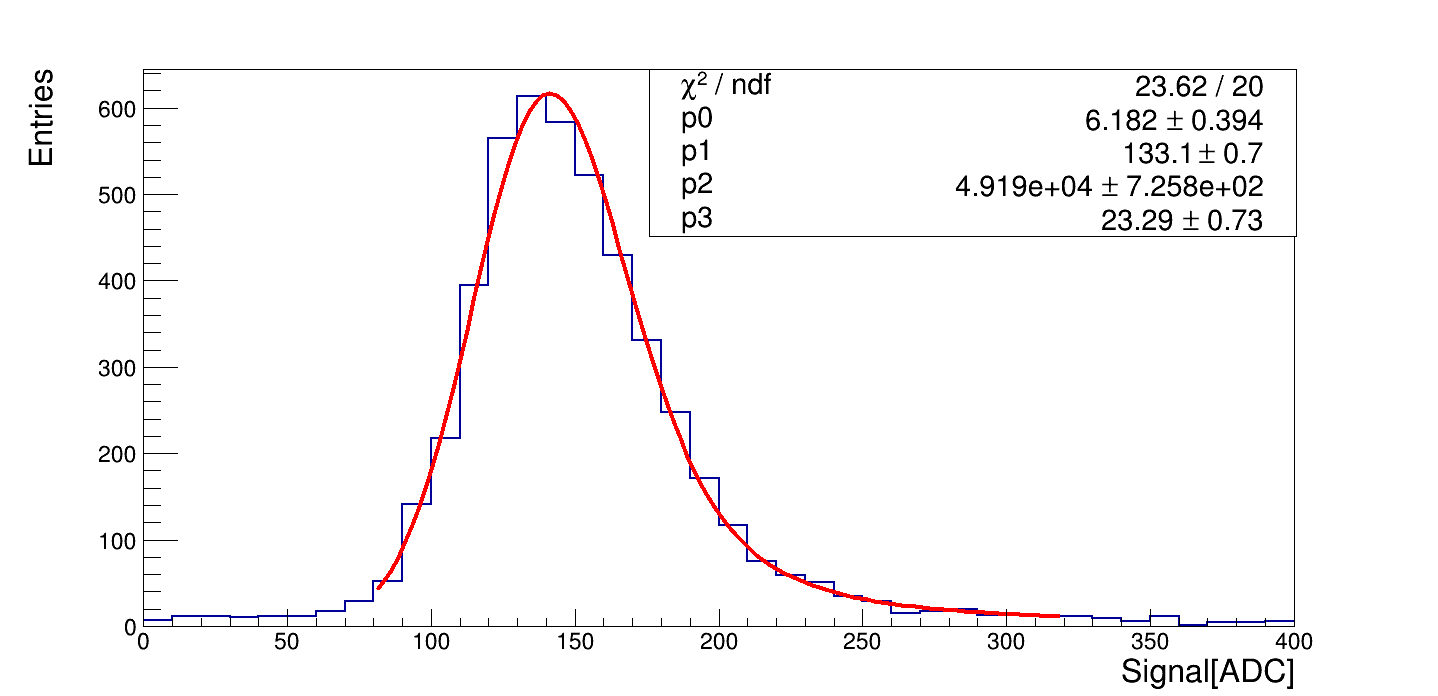}
    \includegraphics[width=0.43\textwidth]{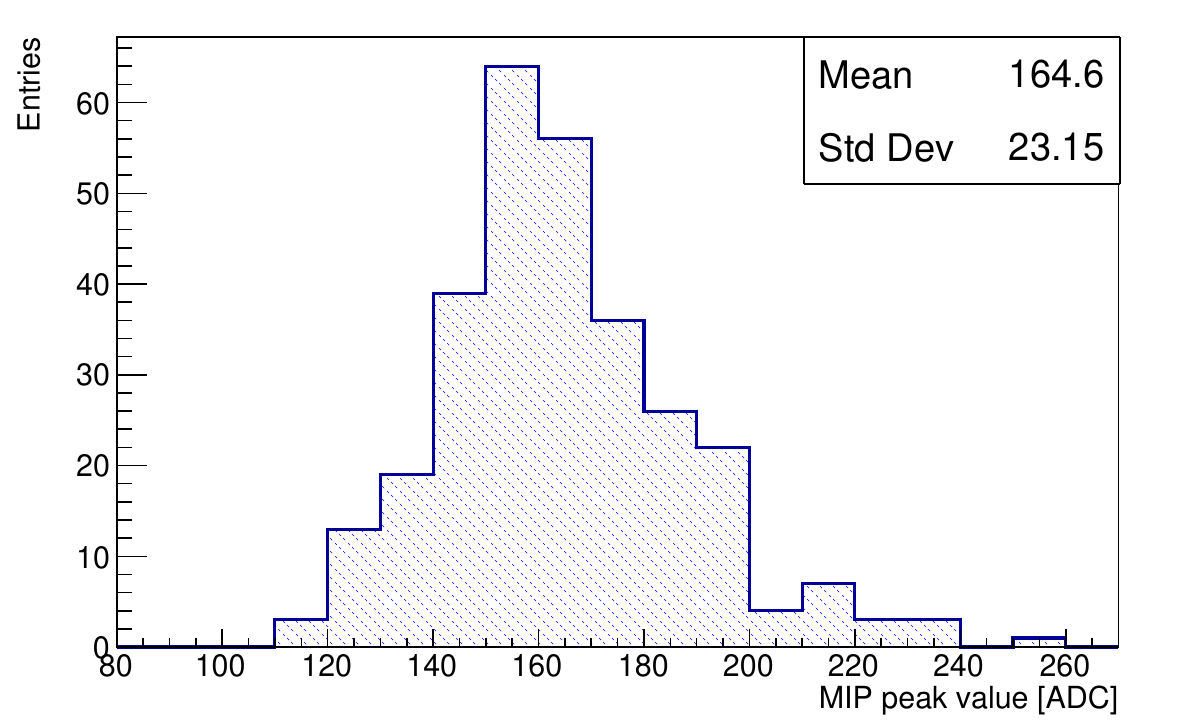}
 	\caption{\label{fig:MIP} Left panel: MIP signals obtained with beam test data for the crystal of tray 3, column 3, layer 3. The histogram is fitted with the convolution of a Gaussian distribution with a Landau distribution. Right panel: histogram with MIP values for all the calibrated channels.}
\end{figure}

Since, for electron beams at the maximum energy of the SPS, the LPDs never reach saturation, only LPDs have been calibrated. SPDs will be calibrated for hadron and nucleus analysis; this procedure will be described in a different paper. The LPDs are calibrated by exploiting the muon energy deposit in LYSO crystals. A scan with a $5\ \text{GeV}$ muon beam of the front face of the calorimeter was performed in order to calibrate all the channels. As a first approximation, $5\ \text{GeV}$ muons are Minimum Ionizing Particles (MIPs); thus, the energy deposits in crystals along the muon track are almost constant. To properly select clean MIP events, by rejecting muons traversing the large amount of CALO passive materials, a simple cut is applied. For each line, the mean of the first and last three cubes are computed. If both values are above given thresholds, the line is considered to be completely traversed by muons, and the event can be used to calibrate all the cubes on the line. The thresholds are selected to be about 3 times the typical noise, which corresponds to about half of the typical MIP signal. The exact values vary slightly across different among different CALO regions to optimize the cut.
Then, we fit the histograms of each cube filled with MIP energy deposit with the convolution of a Gaussian with a Landau distribution. This function is selected to properly take into account the typical distribution of energy deposit in material by MIPs (Landau) and the noise of the electronic system (Gaussian). An example is shown in the left panel of  Figure \ref{fig:MIP}.
We consider the MIP value estimation to be the maximum of the fitted convolution of the Gaussian and Landau distributions. In the right panel of Figure \ref{fig:MIP}, a histogram with the MIP values for all the calibrated channels is shown: the mean MIP signal is about $165\ \text{ADC counts}$, which is very large compared to the mean noise of about $25\ \text{ADC counts}$, resulting in a signal-to-noise ratio better than 6. This value is even better than the one estimated in our previous paper~\cite{bib:HERDpd2022}. The difference can be explained by an improvement in both the gluing and wrapping procedures of the LYSO cubes, which increases the light collected by the PDs.
\begin{figure}[t]
 	\centering 
    \includegraphics[width=0.53\textwidth]{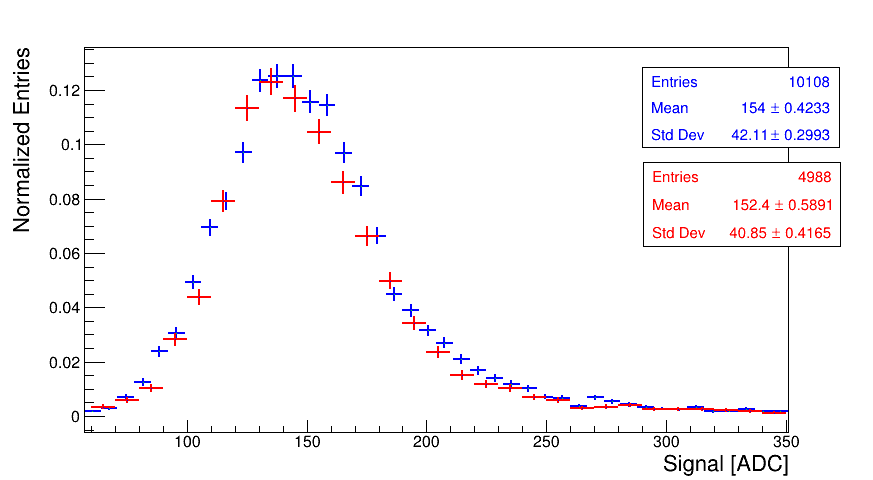}
    \includegraphics[width=0.46\textwidth]{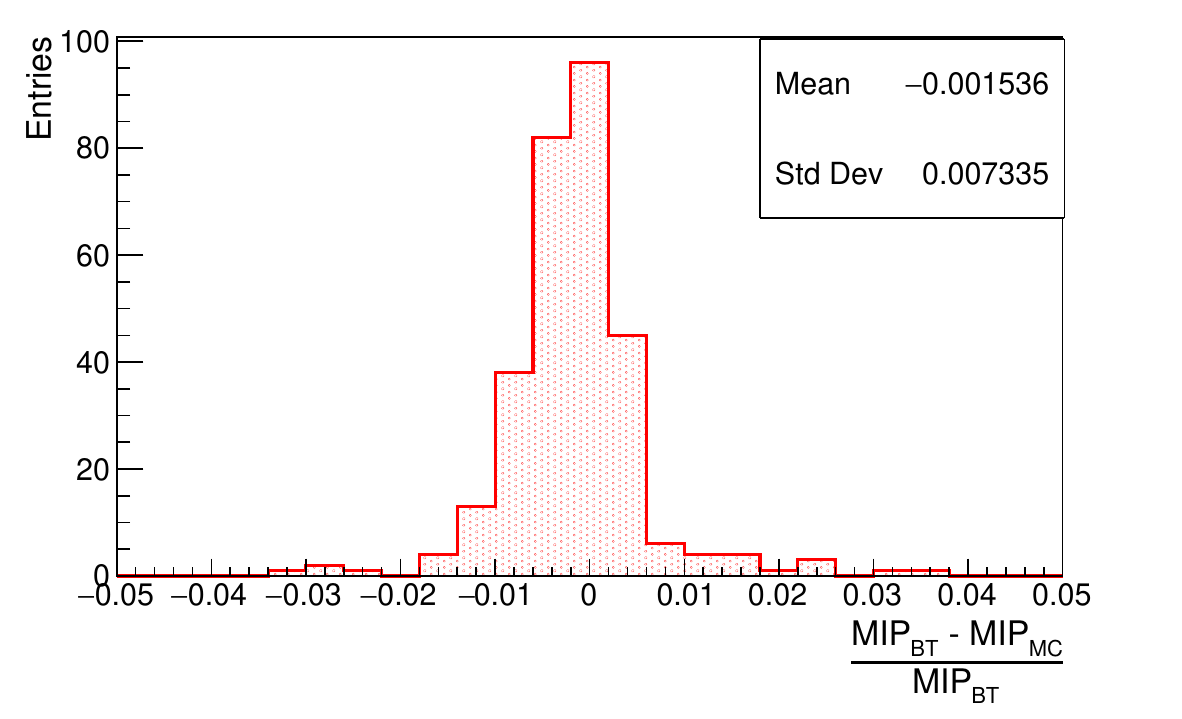}
 	\caption{\label{fig:MCmuon} Left panel: beam test data (red) compared to MC simulation (blue) obtained after the digitization and calibration procedure for the same channel of Figure \ref{fig:MIP}. Right panel: residual difference between MIP peaks estimated with MC and BT data.}
\end{figure}
The next step of the calibration involves Monte Carlo (MC) data and aims to determine the conversion factor between MIP values and electron-volts (eV). As discussed in the previous section, accurately matching the MIP measured with beam test (BT) data requires precise digitization of MC data. However, this process depends on parameters obtained from the calibration itself, such as the MIP-eV conversion factor. Therefore, an iterative process is necessary.\\
Initially, a MIP-eV conversion factor is derived from non-digitized MC simulation. This factor is then used to produce a preliminary approximation of digitized MC data. Comparing the MIP peaks obtained from MC and BT data reveals systematic shifts of approximately 4.5\%. These discrepancies are expected because the initial MIP-eV conversion factor does not account for noise. Using this new version of digitized MC data, we recompute the MIP-eV conversion factors, which now vary among different channels due to their dependence on channel-specific noise. The iterative procedure was repeated three times, as this provided sufficient accuracy in estimating the MIP-eV conversion factors.\\
Figure \ref{fig:MCmuon} presents a comparison between MC and BT data after the iterative procedure. While the distributions do not have exactly the same shape, the agreement in peak position is reasonable. The figure also shows the residual differences between MC and BT peak estimations after the complete calibration-digitization procedure. Only a few channels exhibit differences greater than 2\%.\\
The final step of the calibration is necessary to fully characterize LPD signals when the HiDRA chip operates in low-gain (LG). Typically, muon signals are always read-out with the electronics in high-gain (HG) mode. To account for this, the HG/LG signal ratio for each channel was precisely measured using the HiDRA chip calibration procedure, which was performed several times during the test campaign. As expected, the HG/LG ratio for each channel is approximately 20, with variations below 1\%. This calibration process also confirms the linearity of the chip response, with deviations from ideal linearity being less than 1\%.
By combining the MIP peak positions measured with beam test (BT) data, the MIP-eV conversion factors obtained from MC simulation, and the HG/LG ratios determined through HiDRA chip calibration, the LPD signals are fully calibrated over the entire range, i.e. from few MeV to $\sim 150\ \text{GeV}$ of energy deposited in a single crystal.
\section{Data analysis and high energy electron results.}

The analysis of electron data involves two main steps: (1) the extrapolation of the beam profile and CALO alignment, and (2) the evaluation of CALO performance using five different electron beam energies from both MC and BT data.

\subsection{CALO alignment and beam profile}

\begin{figure}[t]
 	\centering 
    \includegraphics[width=0.55\textwidth]{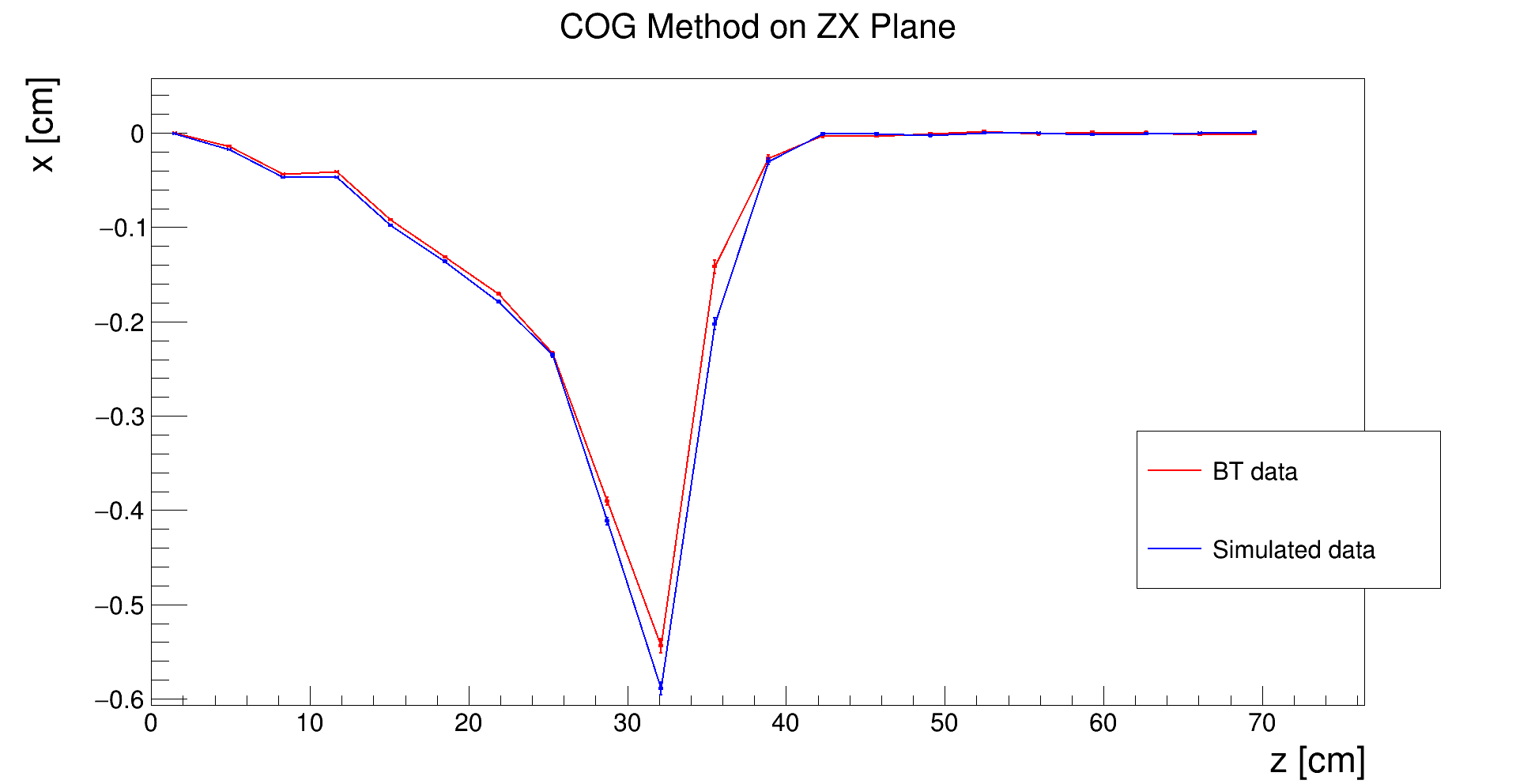}
    \includegraphics[width=0.44\textwidth]{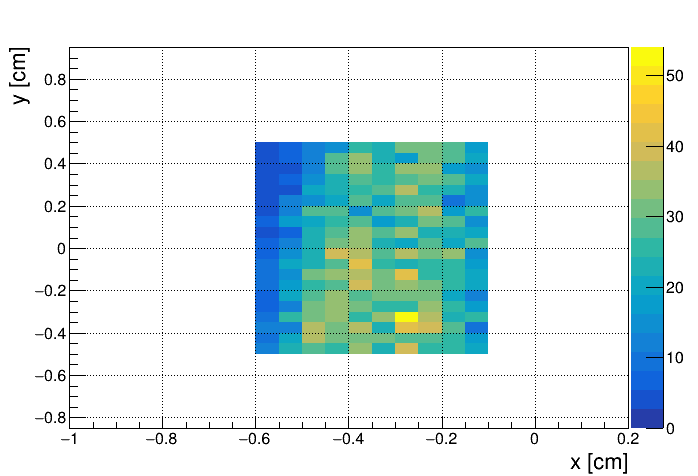}
 	\caption{\label{fig:COG} Left panel: x coordinate of COG as a function of the layer depth obtained with BT (red) and MC (blue) data for $243.48\ \text{GeV}$ electrons. Right panel: profile of $243.48\ \text{GeV}$ beam after the geometrical cut reconstructed with SCD data.}
\end{figure}

The SCD detector~\cite{bib:HERDSCD_2023} consists of a stock of silicon micro-strip detectors, which can measure particle tracks. These tracks were used to extrapolate the beam profile on the entrance face of the CALO and to align the CALO with both the SCD and the beam line. The profile of electron beams was narrower than the dimension of a crystal (i.e., $3\ \text{cm}$) along the x-axis, while it was broader along the y-axis. By selecting events with signals exceeding the equivalent of 20 MIPs in the first crystal of the central line of the CALO, the position of the cube in the SCD reference frame and the beam profile along the y-axis were accurately reconstructed. Unfortunately, this procedure could not be used for the x-axis, as the beam is contained within the central cube, making it difficult to clearly identify the cube's edges and its center position relative to the SCD reference frame.\\
To determine the x-coordinate of the beam entrance point relative to the CALO, a different method was employed. The center of gravity (COG) of each layer was computed using the following equation:
$$
x_{COG} = \frac{1}{E_{layer}} \sum_i E_i \cdot x_i,
$$
where $E_{layer}$ is the total energy deposited in the layer, and $E_i$ and $x_i$ are the energy deposit and the center position of the $i^{th}$ cube, respectively. The COG computed with BT data was compared with those obtained from various MC data samples, each simulating different beam mean entrance points and angles relative to the CALO. This comparison allows the estimation of the beam entrance point and angle that maximize the agreement between MC and BT data. Figure \ref{fig:COG} left panel shows the results for the COG obtained with $200\ \text{GeV}$ BT and MC data. The agreement is reasonable. The trend of the COG as a function of depth is primarily related to the transverse section of the electromagnetic shower, which initially increases as the shower develops, reaches a maximum, and then decreases. Consequently, the COG of central layers is more sensitive to beam alignment than the first and last layers. Therefore, the signals of the last CALO layers are dominated by the noise, so the COG of deep layers is about $0\ \text{cm}$.
\\
After the alignment procedure, well-contained showers in the CALO were selected for the subsequent analysis steps using SCD data. A cut based on the beam entrance point was applied to ensure proper event selection. Figure \ref{fig:COG} (right panel) shows the beam profile within the selected region, which is $0.5 \times 1\ \text{cm}^2$.

\subsection{High energy electrons results}

\begin{figure}[t]
 	\centering 
    \includegraphics[width=0.8\textwidth]{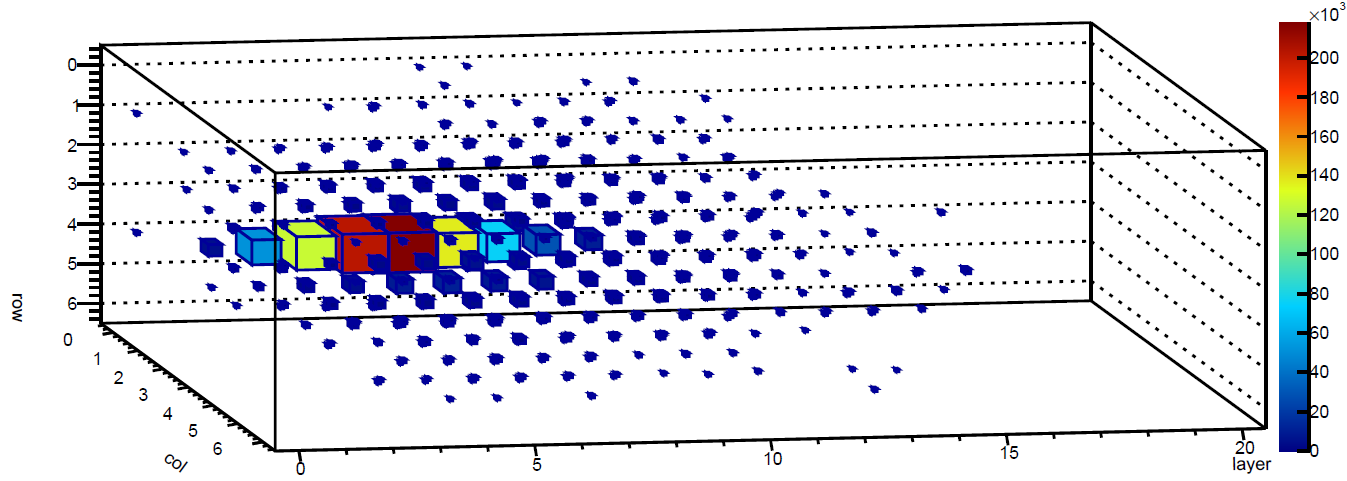}
 	\caption{\label{fig:3Dview} A $243.48\ \text{GeV}$ electron shower acquired with the CALO PD system. For each cube, the color indicates the ADC count of the LPD signal. This image was produced during the online monitoring of the detector, before the signals were calibrated.}
\end{figure}

\begin{figure}[t]
 	\centering 
    \includegraphics[width=0.49\textwidth]{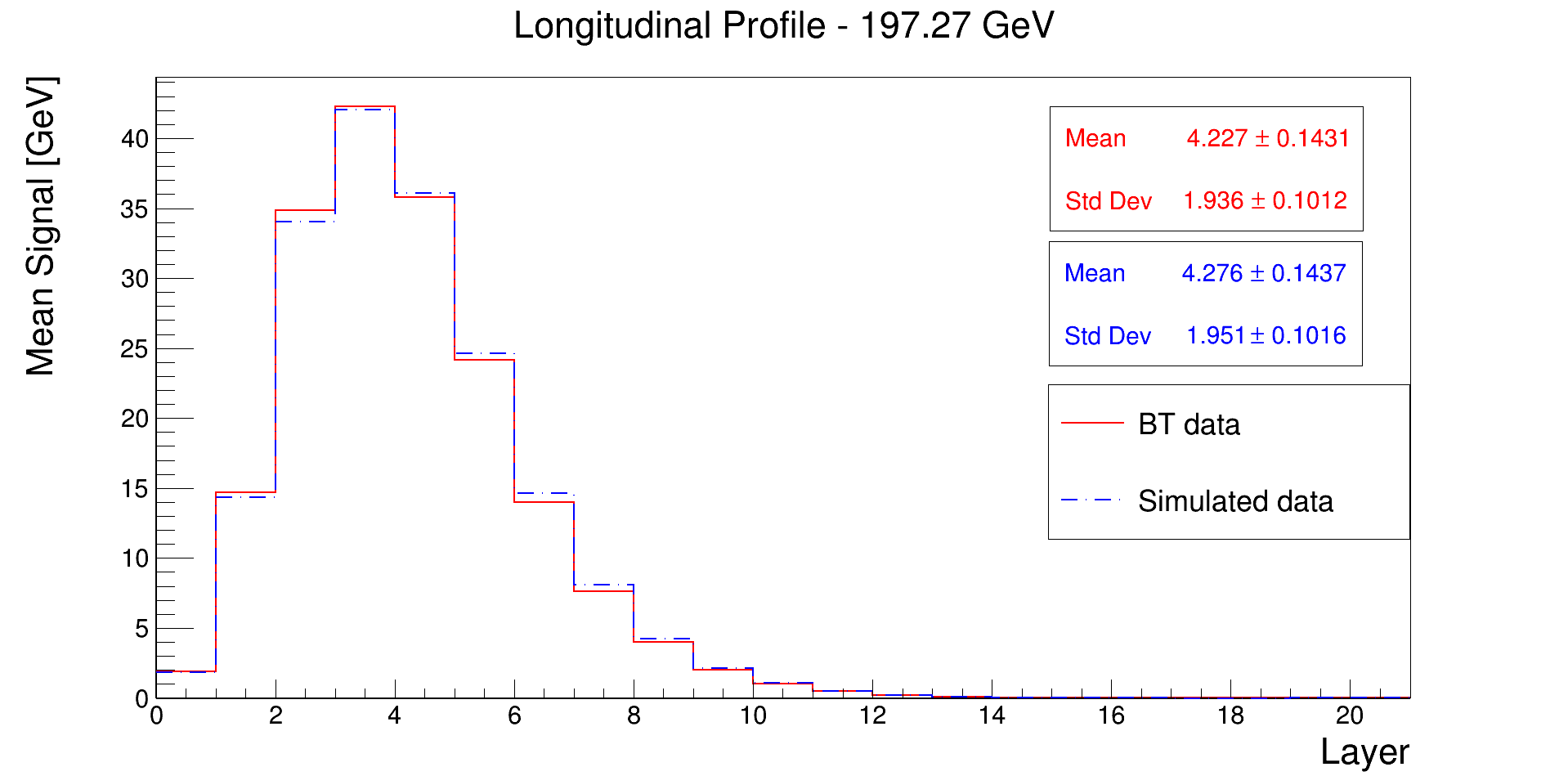}
    \includegraphics[width=0.49\textwidth]{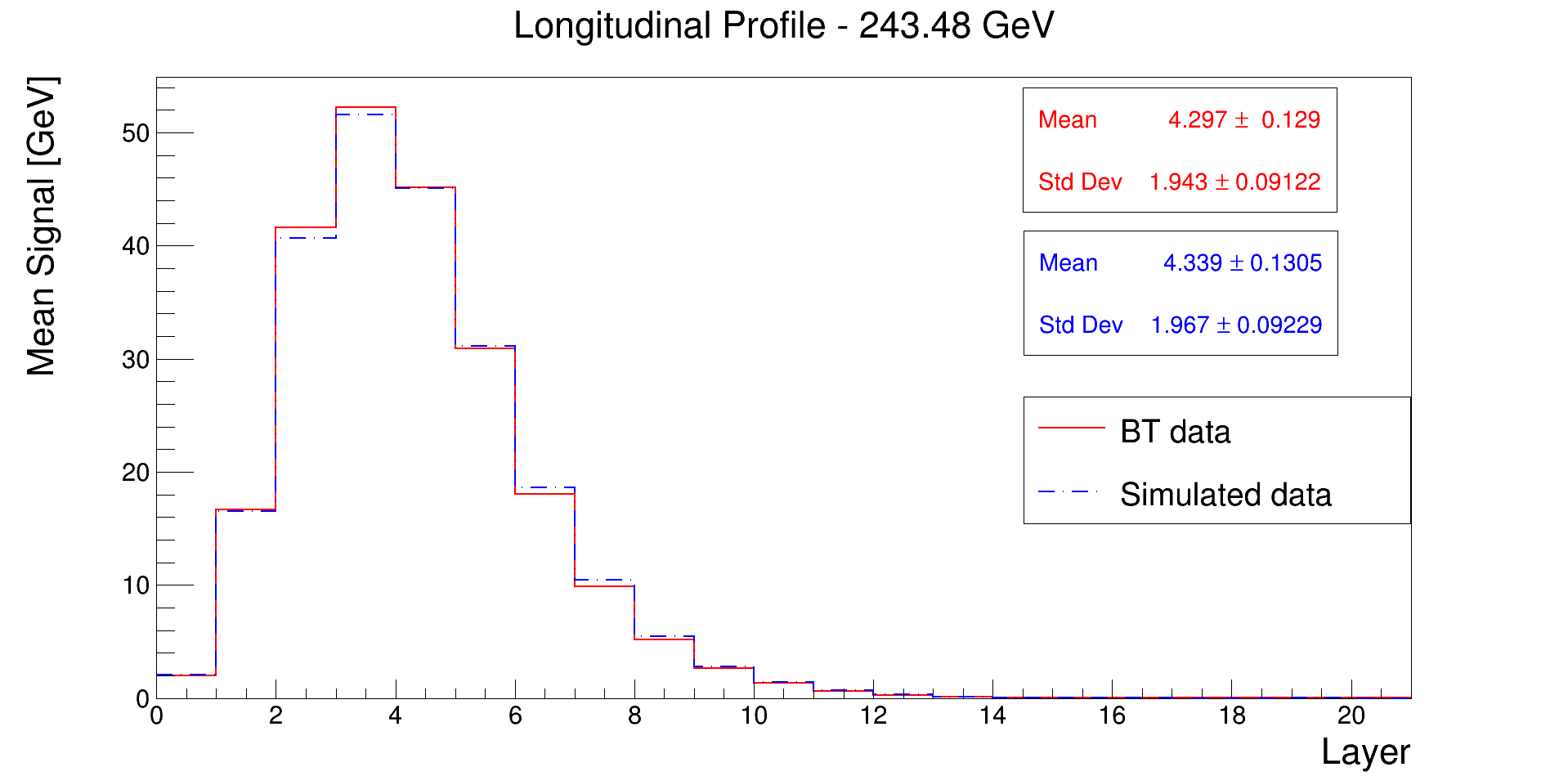}
 	\caption{\label{fig:LongPro} BT (red) and MC (blue) mean longitudinal profile of  $197.27\ \text{GeV}$ and $243.48\ \text{GeV}$ electron showers. The mean energy deposited in each layer is plotted versus the layer number.}
\end{figure}

\begin{figure}[t]
 	\centering 
    \includegraphics[width=0.99\textwidth]{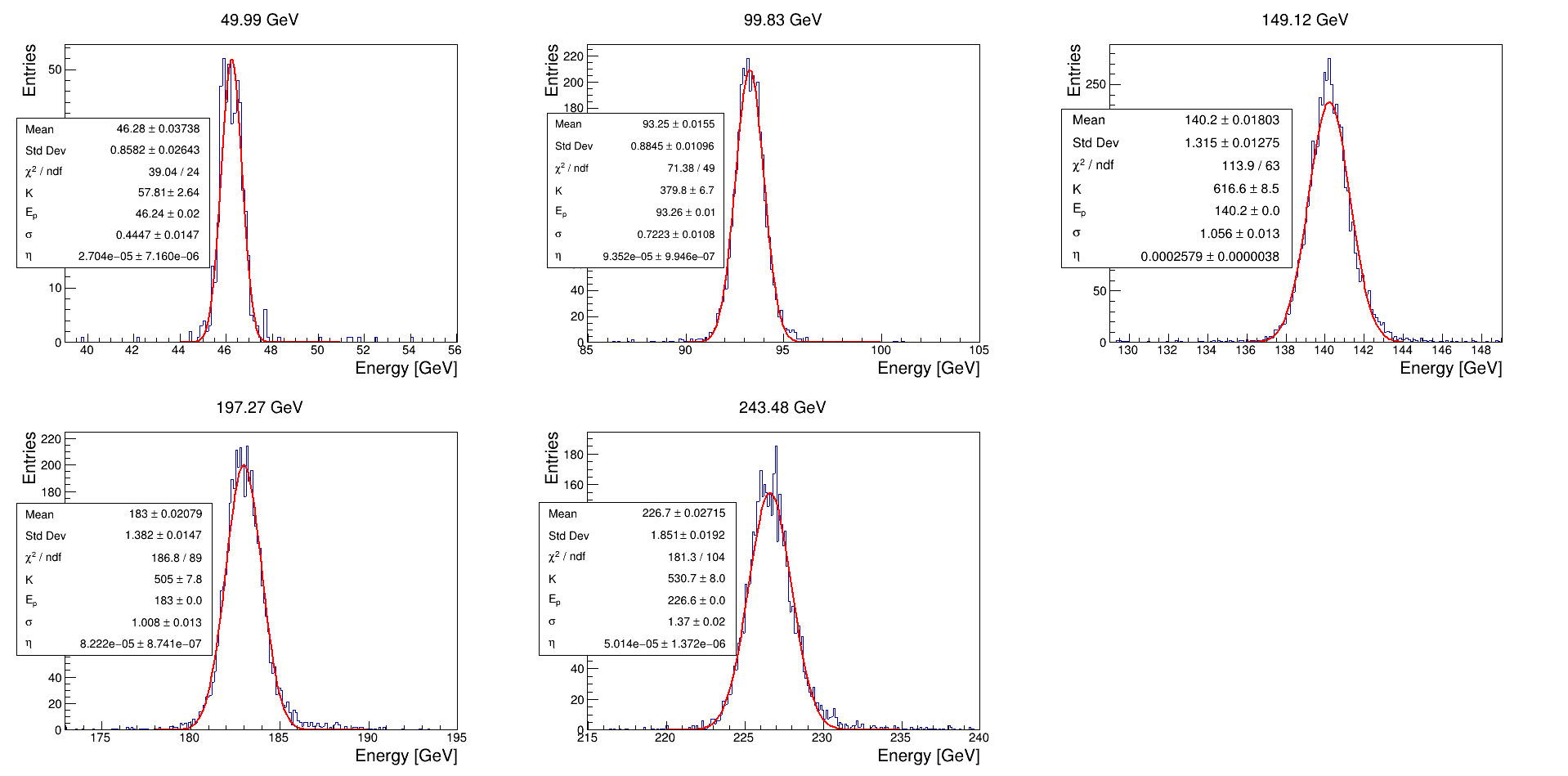}
 	\caption{\label{fig:TotalEDEP} Total energy deposited in CALO for different beam energies (BT data). The red lines are the fit results using equation \ref{eq:logarithmicGaussian}.}
\end{figure}

The detector was able to reconstruct an almost complete 3D image of the electron shower. As an example, Figure \ref{fig:3Dview} shows an event acquired with a $243.48\ \text{GeV}$ electron beam, where the typical shape of an electromagnetic shower in a segmented calorimeter is clearly visible.
Furthermore, the reconstructed longitudinal and lateral shower profiles are in reasonable agreement with those obtained from MC simulation. Figure \ref{fig:LongPro} shows the longitudinal profile obtained with $197.27\ \text{GeV}$ and $243.48\ \text{GeV}$ electron beams. In the MC data, the shower develops slightly deeper compared to the BT data. This discrepancy is likely due to the incomplete accuracy in the simulation of materials in front of the CALO, as discussed in Section \ref{sec:Prototype}.\\
In the final analysis step, for each beam energy, the distribution of the total energy deposited in the CALO is displayed. As explained in \cite{bib:BELLECALO}, to accurately determine the peak of the distributions, these are then fitted with Logarithmic Gaussian distributions, defined as:
\begin{equation}
\label{eq:logarithmicGaussian}
f(E) = K \cdot \frac{\eta}{\sqrt{2 \pi} \sigma \cdot s_{0}} \cdot \exp  \left\lbrace - \frac{\ln^{2} \left[ 1 -\eta ( E - E_{p} ) \right]}{2 s_{0}^{2}} - \frac{s_{0}^{2}}{2} \right\rbrace \text{,}
\end{equation}
\begin{equation}
s_{0} = \frac{2}{2.35} \cdot arcsinh \left( \frac{2.35 \cdot \eta}{2} \right)  \text{,}
\end{equation}
where $E_p$ is the peak energy, $\sigma$ is the Full Width at Half Maximum (FWHM) divided by 2.35, $\eta$ is the asymmetry parameter, and $K$ is the normalization factor. Since the distributions are slightly asymmetric, the standard deviation is not used to estimate the energy resolution. Instead, the resolution is computed as the 68\% confidence interval around the peak divided by the peak value. For a normal distribution, this method is equivalent to the standard definition of resolution, where the standard deviation is divided by the mean. The total energy deposits and the fitted functions for the five beam energies are shown in Figure \ref{fig:TotalEDEP}.\\
\begin{figure}[t]
 	\centering 
    \includegraphics[width=0.95\textwidth]{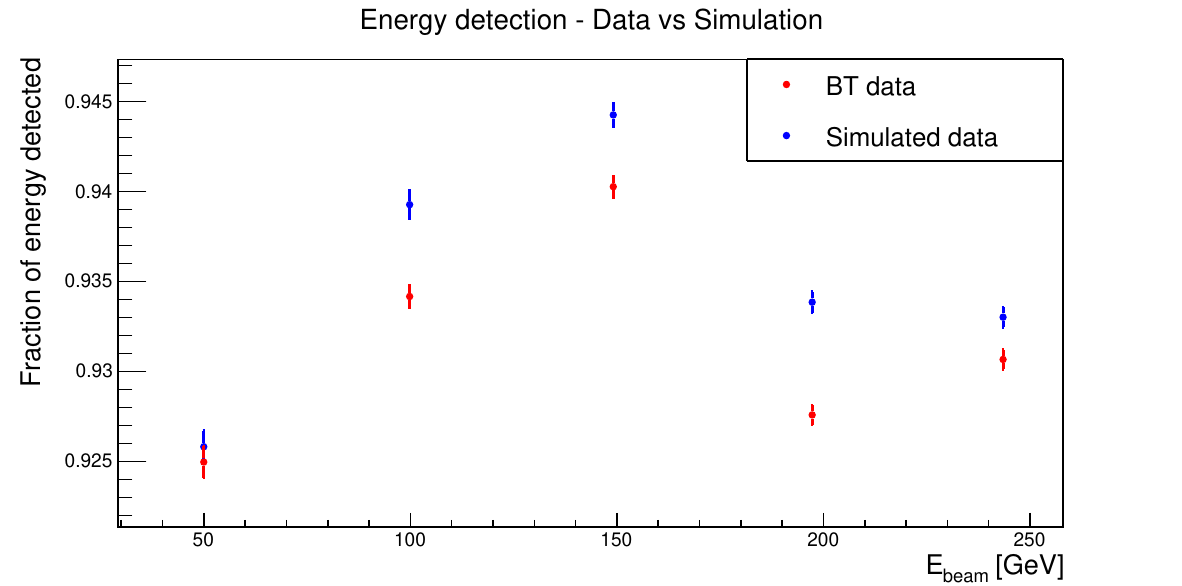}
 	\caption{\label{fig:fraction} Fraction of energy detected for different beam energies, BT data (red) and MC simulation (blue)}
\end{figure}
A key result from this analysis is the fraction of energy deposited in the active volume of the CALO, defined here as the peak energy derived from the fit (which is closely consistent with the histogram mean) divided by the nominal particle energy (i.e., the beam energy). This is illustrated in Figure \ref{fig:fraction}. The energy fraction obtained from BT data ranges between approximately $92.5\%$ and $94.0\%$, while the fraction obtained from MC simulation is slightly higher. The maximum difference is below 1\%, which is considered a reasonable systematic error in the energy scale estimation for such a calorimetric experiment. This discrepancy could be further reduced in the future by implementing more accurate sub-detector models in the simulation, by improving the digitization algorithm, and properly simulating the light yield non-proportionality in LYSO. The impact of the latter on space-borne calorimetric experiments is well described in \cite{bib:NonLinearityScintillator2022}. Apart the absolute values, the trend of the energy fraction as a function of beam energy is well reproduced in the MC simulation. This trend is due to the slightly different longitudinal profiles of the shower at different energies, which affect the fraction of energy deposited in passive materials. Nevertheless, the differences among data points are below 2\%.\\
\begin{figure}[t]
 	\centering 
    \includegraphics[width=0.95\textwidth]{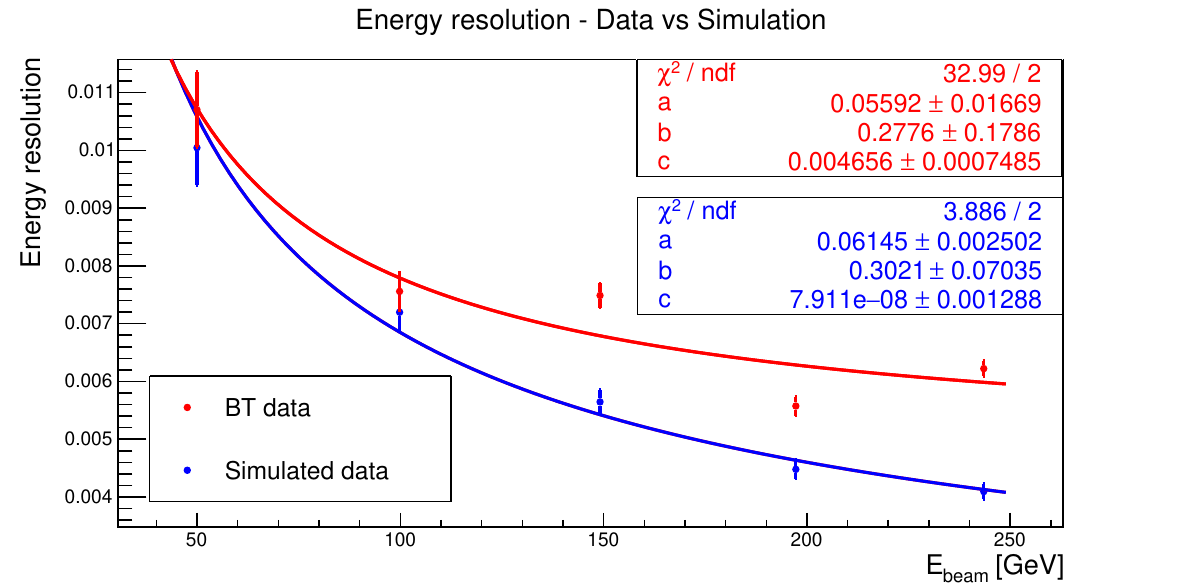}
 	\caption{\label{fig:resolution} Energy resolution obtained with BT (red) and MC (blue) data. Red and blue line are the fit results using equation \ref{eq:enres}.}
\end{figure}
Finally, Figure \ref{fig:resolution} shows the main result of this paper, which is the energy resolution. The red and blue lines represent the fit results for BT and MC data, respectively. The fitted function is the standard one used for calorimeter energy resolution and it is defined as:
\begin{equation}
\label{eq:enres}
\frac{\sigma _{E}}{E} = \sqrt{ \left( \frac{a}{\sqrt{E}} \right) ^{2} + \left( \frac{b}{E} \right) ^{2} + c^{2} }
\end{equation}
where the first term is dominated by statistical fluctuations in the shower development, the second represents the electronic noise component, and the last one accounts for calibration uncertainty and response non-uniformity.\\
The energy resolution estimated from BT data (red points) is excellent: it is approximately $1.1\%$ at $49.99\ \text{GeV}$ and decreases to $0.6\%$ at $243.48\ \text{GeV}$. The fit (red line) is provided as a rough approximation, as it does not perfectly reproduce the measurements. However, the absolute differences between the data points and the fitted function are very small, on the order of $0.1\%$. Given this minimal discrepancy, we did not pursue further investigation. It is reasonable to accept a systematic error of a few per mill in the energy resolution estimation. Moreover, the energy resolution obtained from MC simulations is slightly better than that obtained from BT data. The fit parameters indicate that the calibration term $(c)$ is smaller in the MC simulation, while the other parameters are consistent with the BT results. This is expected, as the calibration procedure for BT data likely introduces systematic errors that are not yet accounted for in the MC simulation. Additionally, as mentioned earlier, several potential improvements to the simulation and digitization processes could enhance the agreement between MC and BT results. Nevertheless, it is reasonable to accept such a systematic error in the energy resolution estimation.

\section{Conclusion.}

In this paper, we present the performance of the HERD calorimeter read-out by the photo-diode system for electron beams with energies ranging from $\sim50\ \text{GeV}$ to $\sim250\ \text{GeV}$. A large prototype was assembled and tested during extensive test campaigns at the PS and NA-SPS. Thanks to a sophisticated calibration procedure that utilizes both PS muon data and simulated data, an accuracy better than a few percent on the main calibration parameters was achieved. Then, to select high-energy electrons impinging on the center of the calorimeter and to fully characterize the beam profile and direction, data from silicon micro-strip detectors were also employed.\\
The analysis results demonstrate that the calorimeter effectively reconstructs the 3D profile of the electromagnetic shower. For example, the longitudinal shower profile is consistent with that obtained from simulations. A large fraction of the primary electron energy, exceeding $90\%$, is detected in the LYSO crystals. Additionally, the detector achieves an excellent energy resolution of approximately $1.1\%$ at $49.99\ \text{GeV}$ and $0.6\%$ at $243.48\ \text{GeV}$. The simulation results are qualitatively consistent with the beam test data, with minor residual discrepancies of a few percent in the fraction of energy detected and a few per mill in the energy resolution.\\
These results demonstrate the exceptional potential of the HERD calorimeter for electron measurement and suggest that the simulation predictions, including those for the flight model of the calorimeter, can be considered reliable.

\bibliographystyle{unsrtnat}
\bibliography{bibliography.bib}
\end{document}